\documentclass[AMA,STIX2COL,Linenumbersoff]{MRM}
\articletype{Full Paper}
\received{XX XXXX 20XX}
\revised{XX XXXX 20XX}
\accepted{XX XXXX 20XX}
\usepackage{graphicx}
\usepackage{amsmath}
\usepackage{amssymb}
\usepackage{array}
\usepackage{makecell}
\usepackage{subcaption}
\usepackage{xcolor}
\usepackage{soul}

\begin{document}

\title{Quiet, rapid 3D multiparametric mapping using magnetization-prepared zero echo time MRI}

\author[1]{Alireza Samadifardheris}{\orcid{0009-0005-3098-6469}}
\author[1]{Shishuai Wang}{\orcid{0009-0008-8550-6591}}
\author[2]{Ana Beatriz Solana}{\orcid{0000-0002-7390-0101}}
\author[3]{Jose de Arcos}{\orcid{0000-0002-1855-4285}}
\author[1]{Noemi Sgambelluri}{}
\author[4,5]{Emil Langensee Ljungberg}{\orcid{0000-0003-1456-7967}}
\author[6]{Sagar Mandava}{\orcid{0000-0003-3425-7579}}
\author[2]{Mika Vogel}{\orcid{0009-0008-3512-7470}}
\author[5]{Steven C. R. Williams}{\orcid{0000-0003-4299-1941}}
\author[1]{Stefan Klein}{\orcid{0000-0003-4449-6784}}
\author[1,7]{Juan Antonio Hernandez-Tamames}{\orcid{0000-0003-0027-9518}}
\author[1]{Dirk H. J. Poot}{\orcid{0000-0003-0656-2963}}
\author[2,5]{Florian Wiesinger}{}

\authormark{SAMADIFARDHERIS \textsc{et al}}

\address[1]{%
  \orgdiv{Department of Radiology and Nuclear Medicine},
  \orgname{Erasmus MC},
  \orgaddress{\city{Rotterdam}, \country{The Netherlands}}}

\address[2]{%
  \orgname{GE HealthCare},
  \orgaddress{\city{Munich}, \country{Germany}}}

\address[3]{%
  \orgname{GE HealthCare},
  \orgaddress{\city{Madrid}, \country{Spain}}}

\address[4]{%
  \orgdiv{Department of Medical Radiation Physics},
  \orgname{Lund University},
  \orgaddress{\city{Lund}, \country{Sweden}}}

\address[5]{%
  \orgdiv{Department of Neuroimaging, Institute of Psychiatry,
  Psychology \& Neuroscience},
  \orgname{King's College London},
  \orgaddress{\city{London}, \country{United Kingdom}}}

\address[6]{%
  \orgname{GE HealthCare},
  \orgaddress{\city{Atlanta}, \country{USA}}}

\address[7]{%
  \orgdiv{Department of Imaging Physics},
  \orgname{TU Delft},
  \orgaddress{\city{Delft}, \country{The Netherlands}}}

\finfo{This work was conducted within the ``Trustworthy AI for MRI'' ICAI
laboratory as part of the ROBUST project, funded by the Dutch Research
Council, GE HealthCare, and the Dutch Ministry of Economic Affairs and
Climate Policy.}

\abstract[Summary]{
\setlength{\baselineskip}{12pt}
\section{Purpose}
To introduce and evaluate MuPa-ZTE, a quiet, rapid 3D framework combining native and magnetization-prepared zero echo time (ZTE) acquisitions for multiparametric mapping.

\section{Methods}
MuPa-ZTE combines steady-state native ZTE with transient-state magnetization-prepared ZTE.
Two implementations were evaluated: T2T1-ZTE for apparent proton density, T1, and T2 mapping, and T1-ZTE for apparent proton density and T1 mapping.
Both were assessed in an ISMRM/NIST system phantom and two healthy volunteers; T2T1-ZTE was also demonstrated in a patient with brain metastases.
The 10-minute phantom and 4.5-minute in vivo acquisitions were retrospectively truncated to 3, 2, and 1 minute and reconstructed with and without deep learning-based denoising.
Evaluations included phantom agreement, precision, short-term repeatability, apparent SNR, edge sharpness, and consistency with full-duration in vivo maps.

\section{Results}
T1 estimates remained close to nominal phantom values across implementations, durations, and reconstructions.
T2 accuracy was maintained down to 2 minutes over the brain-relevant range, with limited sensitivity to longer T2 values.
Denoising generally reduced variability and improved short-term repeatability.
In vivo, 1.1-mm isotropic whole-brain maps were obtained in 4.5 minutes; denoising increased apparent SNR while preserving edge sharpness and yielded promising image quality after retrospective truncation to 2 minutes.

\section{Conclusion}
MuPa-ZTE enables quiet, isotropic 3D multiparametric mapping within 4.5 minutes, supporting robust T1 mapping and T2 mapping over a brain-relevant range, with promising acceleration toward 2 minutes using deep learning-based denoising.
}


\jnlcitation{\cname{%
\author{Samadifardheris A},
\author{Wang S},
\author{Solana AB},
\author{de Arcos J},
\author{Sgambelluri N},
\author{Ljungberg E},
\author{Mandava S},
\author{Vogel M},
\author{Williams SCR},
\author{Klein S},
\author{Hernandez-Tamames JA},
\author{Poot DHJ},
\author{Wiesinger F}} (\cyear{20XX}),
\ctitle{MuPa-ZTE: Quiet, rapid 3D multiparametric mapping using
magnetization-prepared zero echo time MRI}, \cjournal{Magn. Reson. Med.},
\cvol{20XX;XX:X--XX}.}

\maketitle

\footnotetext{Alireza Samadifardheris and Shishuai Wang contributed equally to this work.}
\clearpage

\section{Introduction}
\label{sec:introduction}
Zero echo time (ZTE) MRI is a 3D center-out radial technique in which spatial
encoding begins immediately after RF excitation, enabling acquisition of the
free induction decay with a nominal echo time of zero. The principle was
introduced by Madio and Lowe as the rotating ultrafast imaging sequence
(RUFIS) in the 1990s and revived for modern clinical MRI in the early 2010s
through PETRA and ZTE.\cite{bib_b1,bib_b2,bib_b3}
Excitation occurs with the readout gradient
already at or near its imaging amplitude, eliminating gradient ramping between
excitation and signal acquisition. Without additional gradient activity for
slice selection, prephasing, phase encoding, rewinding, or spoiling, ZTE
enables short repetition times and a high acquisition duty
cycle.\cite{bib_b3,bib_b4,bib_b5} Small changes in gradient direction between
successive radial spokes also produce substantially less acoustic noise than
conventional sequences employing rapid gradient switching.\cite{bib_b4,bib_b6}
Comprehensive reviews describe the principles and applications of ZTE and
short-T2 imaging.\cite{bib_b4,bib_b5}

These properties make ZTE valuable when conventional MRI is limited by rapid
signal decay, susceptibility effects, motion, or acoustic noise. Applications
include cortical-bone and radiation-free CT-like imaging, musculoskeletal
imaging, and free-breathing lung
imaging.\cite{bib_b7,bib_b8,bib_b9,bib_b10,bib_b11,bib_b12,bib_b13,bib_b14} 
Sensitivity to rapidly relaxing signals has also enabled investigation of tissue constituents including the myelin lipid bilayer and collagen.\cite{bib_b15,bib_b16} 
In neuroimaging, T2-prepared RUFIS and Looping Star have enabled quiet,
distortion-reduced whole-brain functional imaging in cognitive,
resting-state, and auditory
paradigms.\cite{bib_b17,bib_b18,bib_b19,bib_b20,bib_b21} Looping Star extends
ZTE through time-multiplexed gradient refocusing, providing an initial
free-induction-decay image followed by gradient echoes with T2* and
susceptibility sensitivity.\cite{bib_b18,bib_b22} Motion-corrected ZTE has
further demonstrated quiet structural neuroimaging with retrospective motion
correction.\cite{bib_b23} Together, these developments establish ZTE as a
versatile platform for structural, functional, and quantitative imaging.

Despite these advantages, native low-flip-angle ZTE provides a restricted range of soft-tissue contrasts.
At the short repetition times and low flip angles typically used, the signal is predominantly proton-density weighted, with residual T1-dependent saturation.\cite{bib_b24}
Magnetization preparation can extend ZTE beyond this native contrast by sensitizing the longitudinal magnetization to selected tissue properties before a segmented readout.
This general approach was introduced with snapshot FLASH and subsequently incorporated into multiparametric methods such as 3D-QALAS, which combines T2 preparation, inversion recovery, and segmented 3D spoiled gradient-echo readouts to estimate T1, T2, and proton density within one acquisition.\cite{bib_b25,bib_b26,bib_b27}
The QALAS preparation scheme was subsequently extended to ZTE readouts, with initial implementations of magnetization-prepared ZTE parameter mapping presented at ISMRM.\cite{bib_b28,bib_b29,bib_b30,bib_b31}

\textbf{MuPa-ZTE} integrates steady-state native ZTE and transient-state magnetization-prepared ZTE within a common 3D radial acquisition framework.\cite{bib_b30} Native ZTE provides apparent proton-density information and an independent signal reference, while readout segments following T1 or T2 preparation encode relaxation-dependent evolution toward the spoiled gradient-echo steady state. Normalization by native ZTE removes common proton-density and receive-sensitivity scaling while preserving the relative amplitudes of the prepared signals, enabling least-squares matching to a physics-based dictionary. Interleaved spiral phyllotaxis provides matched, near-uniform k-space sampling across contrast states and supports flexible acquisition durations.\cite{bib_b23,bib_b32} MuPa-ZTE thus enables isotropic multiparametric relaxometry while retaining the acquisition efficiency, low acoustic noise, and sensitivity to rapidly decaying signals of ZTE.

The present work provides a comprehensive technical description and systematic evaluation of MuPa-ZTE across acquisition durations. Two implementations were investigated: T2T1-ZTE (with T2 preparation and inversion recovery) for apparent proton density, T1, and T2 mapping, and T1-ZTE (only with inversion recovery) for apparent proton density and T1 mapping. Both were evaluated in a quantitative system phantom and healthy volunteers, and T2T1-ZTE was additionally demonstrated in a patient with brain metastases. Conventional reconstruction was used to establish the underlying acquisition and mapping performance, whereas deep learning-based denoising was evaluated for improving precision and supporting shorter acquisitions.\cite{bib_b33}

\section{Methods}
\label{sec:methods}

\begin{figure*}
    \centering
    \includegraphics[width=\linewidth]{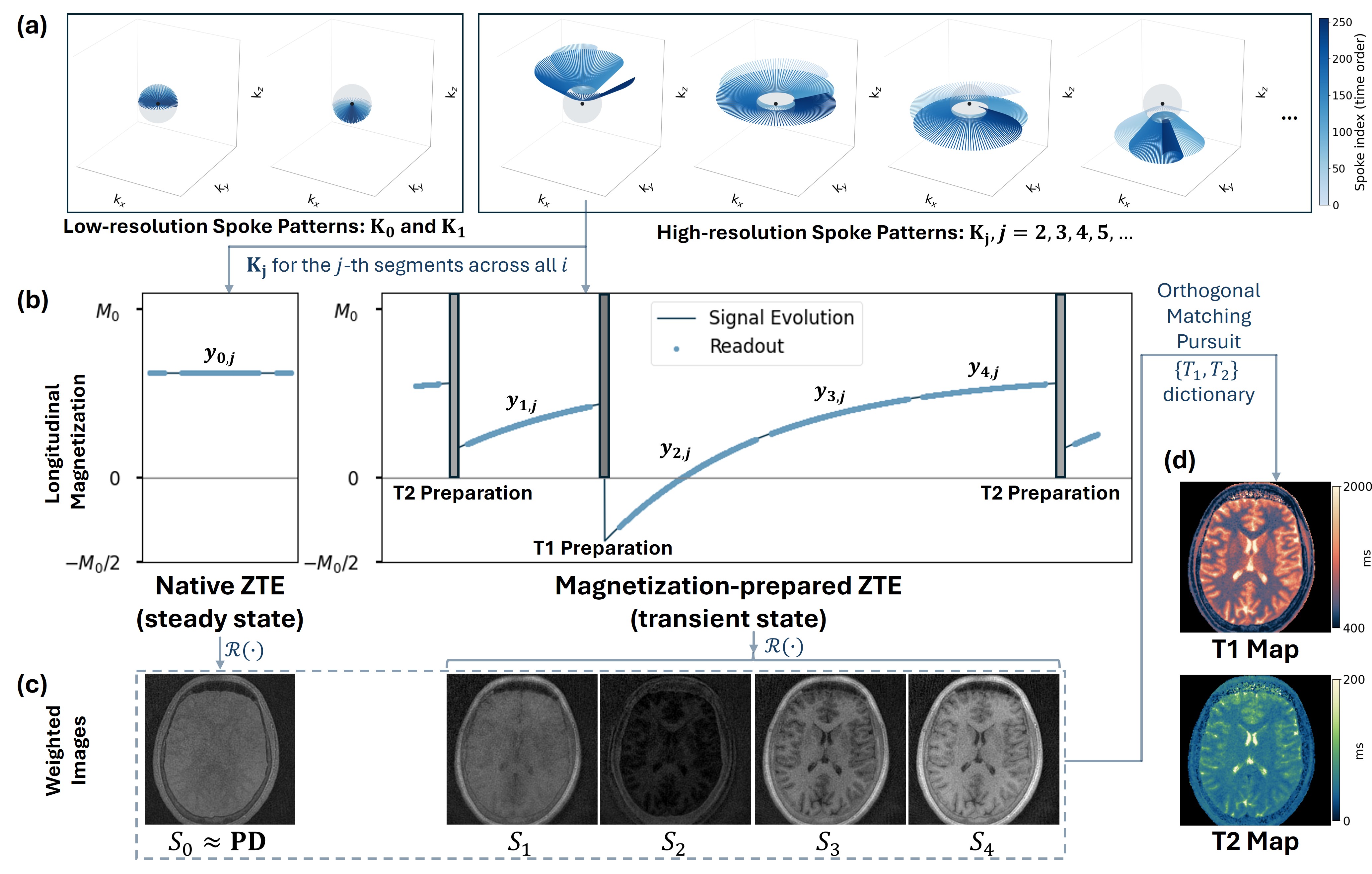}
    \caption{\textbf{MuPa-ZTE acquisition and parameter-mapping workflow,
    illustrated for T2T1-ZTE.} \textbf{A}, Three-dimensional center-out radial
    spokes are ordered using interleaved spiral phyllotaxis patterns. A
    distinct pattern is used for each acquisition repetition and shared across
    contrast states. Two reduced-gradient patterns provide WASPI-type sampling
    of the central k-space region lost during the ZTE transmit-receive dead
    time. \textbf{B}, Native ZTE is combined with magnetization-prepared
    acquisitions. In T2T1-ZTE, T2 preparation precedes $S_1$, followed by
    inversion and acquisition of $S_2$--$S_4$. \textbf{C}, $S_0$ provides
    apparent proton density and normalizes the prepared signals for dictionary 
    matching. \textbf{D}, Representative apparent proton-density, T1, and T2 maps.}
    \label{fig:workflow}
\end{figure*}

Figure~\ref{fig:workflow} summarizes the MuPa-ZTE acquisition,
reconstruction, and parameter-mapping workflow.

\subsection{Signal model}
\label{sec:signalmodel}

MuPa-ZTE uses a segmented ZTE acquisition in which each segment contains $L$
center-out radial spokes. Assuming complete spoiling of transverse
magnetization between excitations (using RF phase cycling and gradient spoiling), the longitudinal magnetization immediately
before excitation $l$ in contrast state $i$ is
\begin{equation}
    \label{eq:mz}
    M_{z,i,l} = M_{z,i,\mathrm{start}}\,\omega_i^{\,l}
    + M_{z,\mathrm{ss},i}\left(1-\omega_i^{\,l}\right),
    \quad l = 0,\ldots,L-1,
\end{equation}
where
\begin{equation*}
    \omega_i = E_1\cos(\alpha_i),
    \qquad
    E_1 = \exp\!\left(-\frac{TR}{T_1}\right),
\end{equation*}
and
\begin{equation*}
    M_{z,\mathrm{ss},i} = PD\,\frac{1-E_1}{1-E_1\cos(\alpha_i)}
\end{equation*}
is the spoiled gradient-echo steady-state longitudinal magnetization for
excitation flip angle $\alpha_i$. $M_{z,i,\mathrm{start}}$ is the longitudinal
magnetization at the beginning of the segment, $TR$ is the interval between
excitations, and $PD$ denotes proton density.

Because every radial spoke samples the center of k-space, source-image
contrast was approximated by the longitudinal magnetization averaged across
the segment:
\begin{equation}
    \label{eq:mzbar}
    \bar{M}_{z,i} = M_{z,i,\mathrm{start}}\,f(L,\omega_i)
    + M_{z,\mathrm{ss},i}\left[1-f(L,\omega_i)\right],
\end{equation}
where
\begin{equation}
    \label{eq:f}
    f(L,\omega_i) = \frac{1-\omega_i^{\,L}}{L\left(1-\omega_i\right)}.
\end{equation}

After each segment, the magnetization was propagated through the measured
delay by T1 relaxation and modified by the subsequent preparation module. The
dictionary therefore incorporated the complete preparation and readout
history.

Neglecting transverse decay and off-resonance during the short ZTE readout,
the signed source-image signal was modeled as
\begin{equation}
    \label{eq:signal}
    S_i(\mathbf{r}) = C(\mathbf{r})\sin(\alpha_i)\,\bar{M}_{z,i}(\mathbf{r}),
\end{equation}
where $C(\mathbf{r})$ represents receive-coil sensitivity and
system-dependent scaling.
The model represents magnetization evolution across the $L$ spokes by one
effective signal per source image and hence ignores k-space modulation due to longitudinal magnetization evolution during a segment. 

\subsection{MuPa-ZTE acquisition}
\label{sec:acquisition}

\subsubsection{Three-dimensional radial sampling}
\label{sec:sampling}

Three-dimensional center-out radial spokes were ordered using interleaved
spiral phyllotaxis patterns.\cite{bib_b23,bib_b32} A distinct pattern was used
for each acquisition repetition, and the same pattern was shared across all
contrast states within that repetition. This matched sampling minimizes
contrast-dependent differences in k-space coverage and permits acquisition
shortening by retaining complete repetitions. The trajectory is also
compatible with self-navigation and retrospective motion correction, although
motion correction was not applied in the present study.\cite{bib_b23}

The central k-space region lost during the finite transmit-receive dead time
was sampled using a WASPI-type strategy.\cite{bib_b34,bib_b35} Two additional
patterns were acquired with an eightfold-reduced readout-gradient
amplitude, thereby reducing the radius of the missing region. Full-gradient
strength patterns provided the nominal spatial resolution.

\subsubsection{MuPa-ZTE framework}
\label{sec:framework}

MuPa-ZTE combines two complementary signal regimes:
1) \textbf{steady-state native ZTE}, providing the apparent proton-density reference $S_0$; and
2) \textbf{transient-state magnetization-prepared ZTE}, providing images $S_1,\ldots,S_4$.
Native ZTE was acquired without magnetization preparation using a low flip
angle $\alpha_0$. Prepared source images were acquired with flip angle
$\alpha$ after adiabatic T1 or T2 preparation to reduce sensitivity to
transmit-field inhomogeneity. Inversion used a $180^{\circ}$ adiabatic pulse,
and T2 preparation used a $90^{\circ}$--$180^{\circ}$--$180^{\circ}$--$90^{\circ}$ train of BIR-4 pulses.\cite{bib_b36,bib_b37,bib_b38}

Two implementations were evaluated:
\begin{itemize}
    \label{eq:variants}
    \item \textbf{T2T1-ZTE}: T2 Prep.-$S_1$–Inversion–$S_2$–$S_3$–$S_4$
    \item \textbf{T1-ZTE}: Inversion–$S_1$–$S_2$–$S_3$–$S_4$
\end{itemize}

T2T1-ZTE supports apparent proton density, T1, and T2 mapping. T1-ZTE omits T2
preparation and supports apparent proton density and T1 mapping. Because $S_0$
is common to both implementations, it is omitted from their names.

No dedicated recovery interval was inserted between segments. Consecutive
source images therefore sample the continued longitudinal-magnetization
evolution rather than independent fully recovered states.

The source images also provide complementary contrasts. Intensity inversion of
$S_0$ provides CT-like visualization of bone.\cite{bib_b7} Post-inversion
images provide inversion-recovery anatomical contrast and can generate
inflow-sensitive angiographic contrast using minimum-intensity projection.

\subsection{Reconstruction and parameter mapping}
\label{sec:recon}

\subsubsection{Image reconstruction}
\label{sec:imagerecon}

Data from each receive-coil channel were reconstructed using
density-compensated 3D Kaiser-Bessel gridding, followed by Fourier
transformation and deapodization.\cite{bib_b36,bib_b37} Coil images were
combined using the same sensitivity-weighted procedure for all contrast
states.

A deep learning based method for ZTE noise and artifact suppression can be added, indicated by \textbf{DL On}.\cite{bib_b33,bib_b38,bib_b39,bib_b40} For ZTE imaging this improved precision and allowed shorter acquisitions.  However, this processing is not intrinsic to MuPa-ZTE and hence with \textbf{DL Off} we evaluate the acquisition and mapping performance with conventional reconstruction.



\subsubsection{Apparent proton-density normalization}
\label{sec:normalization}

The native ZTE image $S_0$ provides an apparent proton-density estimate and serves as the reference for the prepared source images.
The common spatial factors in Eq.~\ref{eq:signal} were removed by normalization to $S_0$.
Mild T1-dependent saturation in $S_0$ was corrected iteratively using the preceding T1 estimate,
as described in Supporting Information~A.

For each voxel, the normalized measurement vector was
\begin{equation}
    \label{eq:normvector}
    \mathbf{S} =
    \begin{bmatrix}
        S_1/S_{0,\mathrm{corr}}\\
        S_2/S_{0,\mathrm{corr}}\\
        S_3/S_{0,\mathrm{corr}}\\
        S_4/S_{0,\mathrm{corr}}
    \end{bmatrix},
\end{equation}
where $S_{0,\mathrm{corr}}$ denotes the saturation-corrected native ZTE
signal. Normalization removes common receive-sensitivity and proton-density
scaling, reducing the parameter search to $(T_1,T_2)$ for T2T1-ZTE and $T_1$
for T1-ZTE.

The corrected native image was retained as apparent proton density. No
absolute proton-density calibration was performed, and the image retains
sequence- and system-dependent scaling and sensitivity to rapidly decaying
components.

\subsubsection{Dictionary matching}
\label{sec:dictionary}

Physics-based dictionary atoms were generated using
Equations~\ref{eq:mz}--\ref{eq:signal}, including the preparation modules,
excitation flip angles, relaxation delays, and magnetization evolution during
the segmented readouts. 
Because there is minimal recovery delay separates preparation cycles, the cycle was repeated 20 times to reach the cyclic steady state. 

For T2T1-ZTE the dictionary $\mathbf{d}(T_1,T_2)$ was created for 1000 T1, linearly spaced between 10 and 4000 ms and 1000 T2 linearly spaced between 5 and 2000 ms, hence consisting of $10^6$ atoms atoms in total. For T1 ZTE the dictionary $\mathbf{d}(T_1)$ consists of 1000 atoms varying in T1.
Matching was performed voxel-wise by exhaustive grid search. 
For T2T1-ZTE,
\begin{equation}
    \label{eq:match_t2t1}
    (\hat{T}_1,\hat{T}_2) =
    \operatorname*{arg\,min}_{T_1,T_2}
    \left\| \mathbf{S} - \mathbf{d}(T_1,T_2) \right\|_2^2 .
\end{equation}
For T1-ZTE,
\begin{equation}
    \label{eq:match_t1}
    \hat{T}_1 =
    \operatorname*{arg\,min}_{T_1}
    \left\| \mathbf{S} - \mathbf{d}(T_1) \right\|_2^2 .
\end{equation}

Least-squares matching evaluates both signal shape and amplitude relative to
$S_0$. In contrast, normalized inner-product matching, commonly used in
magnetic resonance fingerprinting, removes signal scale.\cite{bib_b41,bib_b42} The
native ZTE reference enables MuPa-ZTE to remove common system scaling while
retaining relative-amplitude information.

Effective excitation, inversion, and T2-preparation efficiency factors were
included to account for systematic deviations from the idealized RF model.\cite{bib_b43,bib_b44,bib_b45}
Their phantom-based selection is described in A of the Supporting Information~A.

\subsection{Experimental acquisitions}
\label{sec:experiments}

\subsubsection{Study design and MRI system}
\label{sec:studydesign}

Experiments included an ISMRM/NIST system phantom (Premium System Phantom,
Model 130; CaliberMRI, Boulder, CO, USA), two healthy volunteers, and one
patient with brain metastases.\cite{bib_b46}

All measurements were performed at 3\,T using a SIGNA Premier scanner and a
48-channel brain coil (GE HealthCare, Chicago, IL, USA). Both in vivo
protocols were approved by the Erasmus MC WMO Ethics Committee (Medical
Ethics Review Committee, Erasmus MC, Rotterdam, The Netherlands): healthy
volunteers were scanned under protocol NL47429.078.13 and the patient under
protocol NL86761.078.24. All participants provided written informed consent
before the examination.

\subsubsection{Acquisition parameters}
\label{sec:acqparams}

Both implementations used a $192\times192\times192$\,mm\textsuperscript{3}
field of view, a $176\times176\times176$ acquisition matrix, 1.1-mm isotropic
nominal resolution, $TR = 1.7$\,ms for each spoke, nominal $TE = 0$, and receiver bandwidth
$\pm31.25$\,kHz. Native and prepared ZTE flip angles were $0.9^{\circ}$ and
$3.0^{\circ}$, respectively. Each segment contained 256 spokes. Each
center-out spoke was sampled with 179 readout points at twofold readout
oversampling. Segments were preceded and followed by a 10-ms gradient ramp and
followed by an 8-ms inter-segment interval, and no delay separated successive
preparation cycles.
T2T1-ZTE used an adiabatic BIR-4-based T2-preparation module with an effective preparation
TE of 80\,ms and an adiabatic $180^{\circ}$ inversion pulse. The T2-preparation module had a
total duration of 108.7\,ms, and the inversion pulse was followed by a 30-ms
inversion delay before acquisition of $S_2$. T1-ZTE used the same inversion
preparation without T2 preparation.
The 10-minute phantom acquisition contained 61,952 spokes, including 512
WASPI spokes. The 4.5-minute in vivo acquisition contained 28,160
spokes, including 512 reduced-gradient spokes.

\subsubsection{Phantom and in vivo acquisitions}
\label{sec:acquisitions}

Both MuPa-ZTE implementations were acquired twice consecutively without
repositioning in the phantom and the two healthy volunteers. The 10-minute
phantom datasets were retrospectively truncated to 4.5, 3, 2, and 1 minute.
The 4.5-minute volunteer datasets were truncated to 3, 2, and 1 minute.
Complete acquisition repetitions were retained to preserve preparation order
and matched sampling.

A single 4.5-minute T2T1-ZTE acquisition was obtained in the patient and
retrospectively truncated in the same manner.

\subsection{Phantom analysis}
\label{sec:phantomanalysis}

Sphere masks were defined from the first full-duration T2T1-ZTE DL Off dataset and applied unchanged to all conditions. T1 was evaluated in the NiCl$_2$ spheres (T1 plate) T1-01 to T1-06, spanning nominal values of 351.5--1989\,ms.
T2 was evaluated in the MnCl$_2$ spheres (T2 plate) T2-04 to T2-09, spanning 45.3--267.0\,ms. Each sphere has known reference T1 and T2 values. The phantom was stored in the scanner room and allowed to equilibrate before acquisition;
its measured temperature was 20\,$^{\circ}$C, matching the manufacturer's reference tabulation used for nominal values.

For sphere $i$, absolute error was
\begin{equation}
    \label{eq:abserror}
    \epsilon_i = \left| \hat{p}_i - p_i^{\mathrm{nom}} \right| ,
\end{equation}
and within-sphere standard deviation was
\begin{equation}
    \label{eq:sd}
    \sigma_i = \sqrt{\frac{1}{N_i}\sum_{v \in i}
    \left(p_{i,v}-\hat{p}_i\right)^2 } .
\end{equation}
Short-term repeatability was assessed as
\begin{equation}
    \label{eq:repeat}
    d_i = \left| \hat{p}_{i,\mathrm{rep1}} - \hat{p}_{i,\mathrm{rep2}}
    \right| .
\end{equation}

Because the measurements were consecutive and performed without repositioning,
$d_i$ reflects short-term repeatability rather than intersession
reproducibility.

\subsection{In vivo analysis}
\label{sec:invivoanalysis}

Quantitative analyses were performed for T2T1-ZTE in Volunteer 1.

Apparent SNR was measured in ten $9\times9$-voxel white-matter regions:
\begin{equation}
    \label{eq:snr}
    \mathrm{SNR}_{\mathrm{app}} = \frac{\mu_{\mathrm{ROI}}}
    {\sigma_{\mathrm{ROI}}} .
\end{equation}
Because the spatial standard deviation includes anatomical variation and
potentially correlated noise, the quantity is reported as apparent SNR.

Edge sharpness was evaluated at 100 manually selected tissue boundaries.
Boundary-normal profiles were sampled using linear interpolation and normalized to $[0,1]$. Sharpness was defined as
\begin{equation}
    \label{eq:sharpness}
    \mathcal{S} = \frac{1}{\left| t_{90}-t_{10} \right|} .
\end{equation}

Mean absolute error was calculated within a common head mask:
\begin{equation}
    \label{eq:mae}
    \mathrm{MAE} = \frac{1}{N}\sum_{v=1}^{N}
    \left| p_v - p_{v,\mathrm{ref}} \right| ,
\end{equation}
using the corresponding 4.5-minute DL On map as the reference. MAE therefore
measures consistency with the selected reference rather than absolute
accuracy.

Complementary source-image contrasts and patient maps were evaluated
descriptively. No inferential statistical testing was performed.

\section{Results}
\label{sec:results}

\subsection{Phantom performance}
\label{sec:res_phantom}

\begin{figure*}[!t]
  \centering
  \begin{minipage}[t]{0.48\linewidth}
    \includegraphics[width=\linewidth]{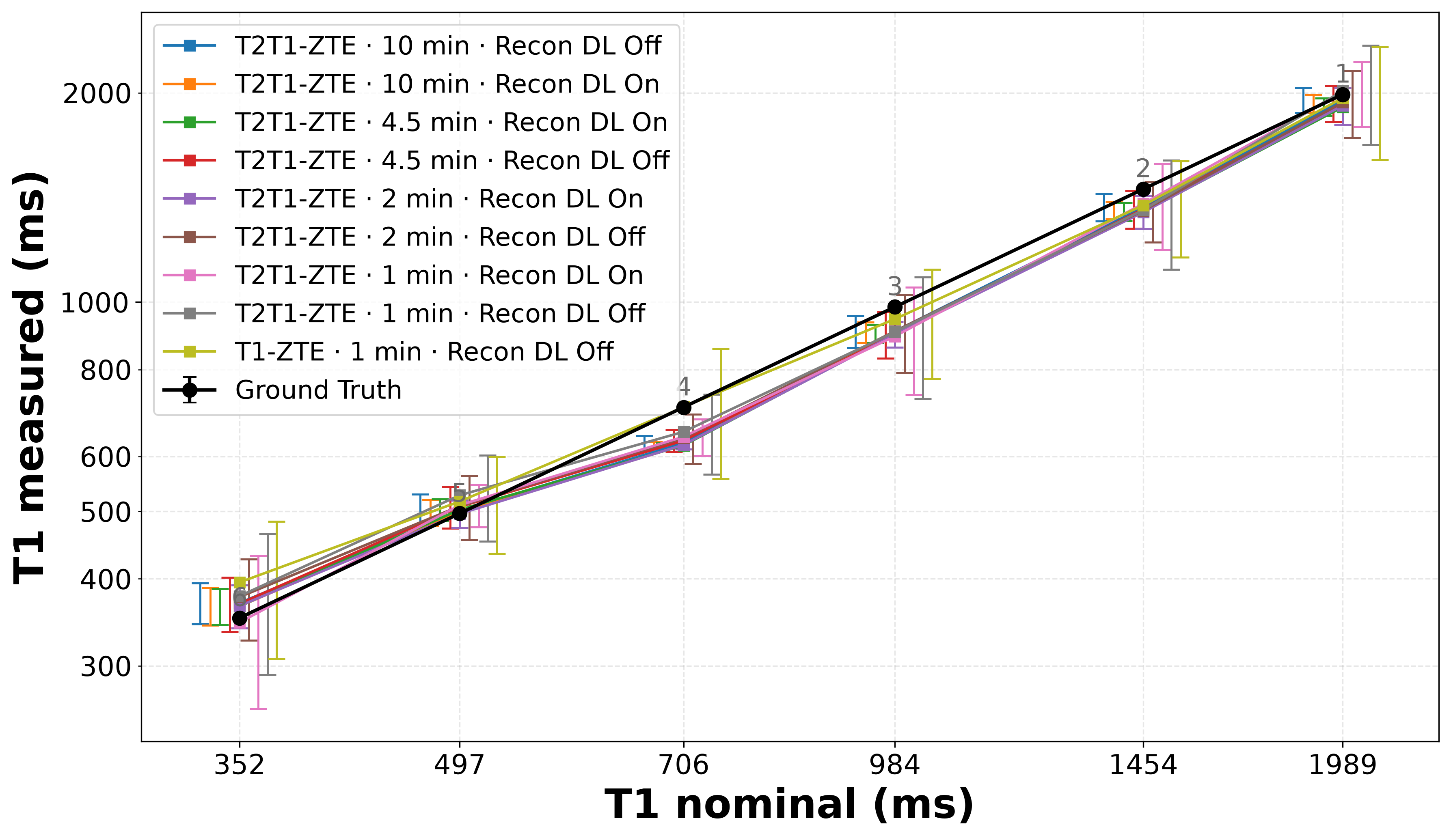}
  \end{minipage}\hfill
  \begin{minipage}[t]{0.48\linewidth}
    \includegraphics[width=\linewidth]{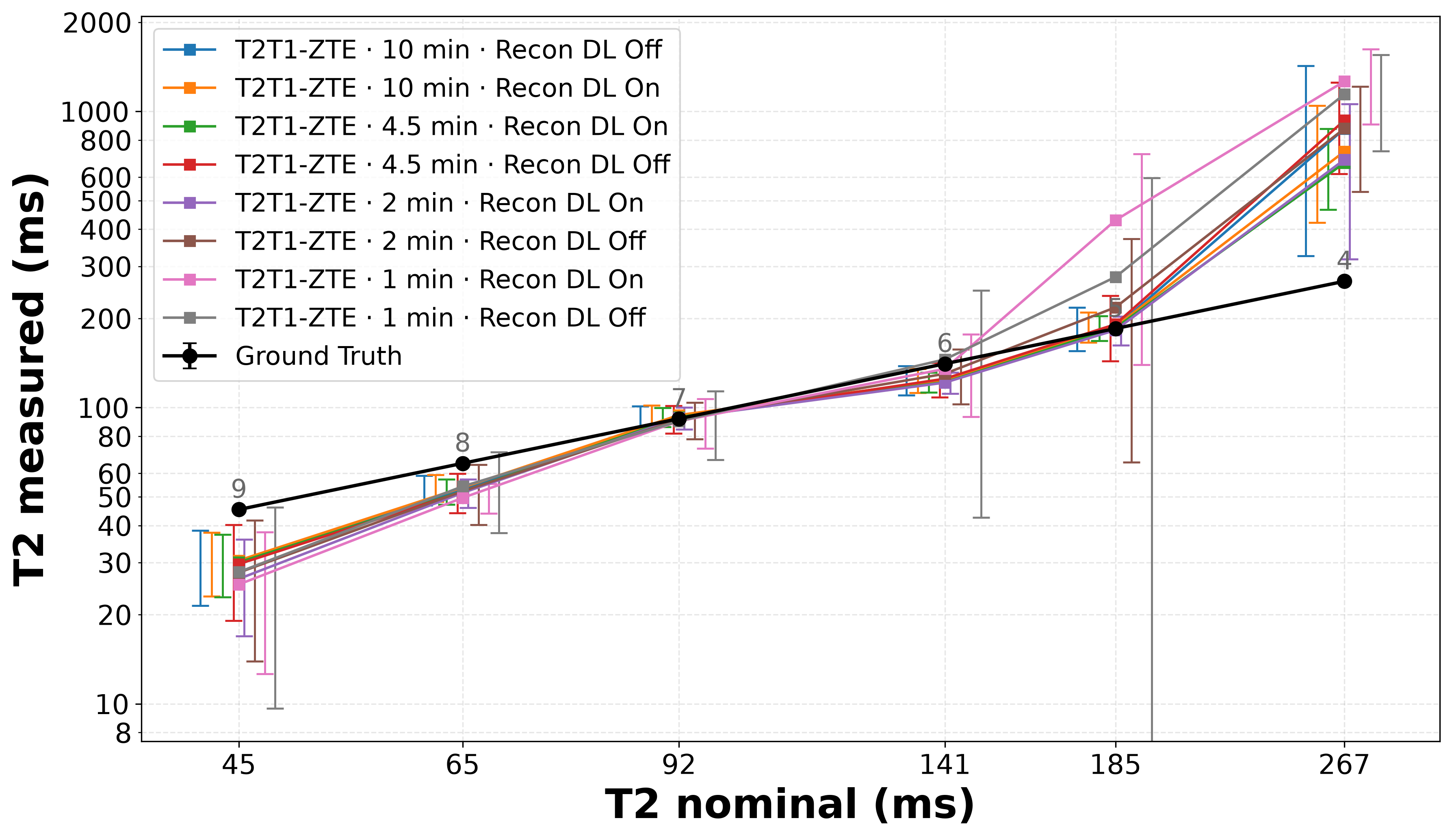}
  \end{minipage}
  \caption{\textbf{Measured versus nominal relaxation times in the ISMRM/NIST
  system phantom.} T1 and T2 estimates are shown for representative
  acquisition durations and reconstruction settings. Markers and error bars
  denote sphere means and within-sphere standard deviations. The black line
  indicates identity.}
  \label{fig:scatter}
\end{figure*}

\begin{figure*}[!t]
    \centering
    \includegraphics[width=\textwidth]{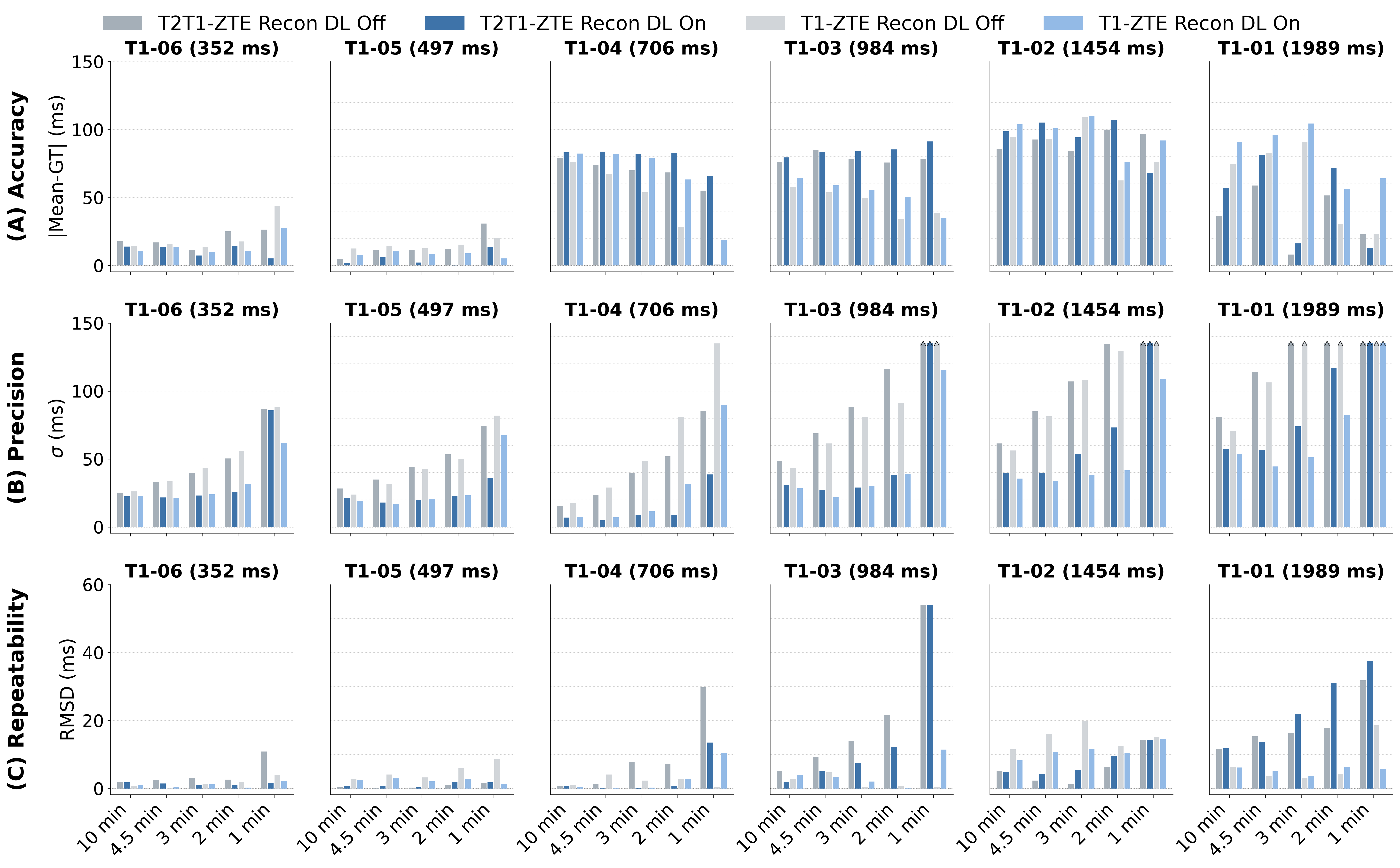}
    \caption{\textbf{T1 accuracy, precision, and short-term repeatability.}
    Sphere-wise \textbf{A}, absolute error relative to nominal T1;
    \textbf{B}, within-sphere standard deviation; and \textbf{C}, absolute
    scan--rescan difference are shown across acquisition durations, MuPa-ZTE
    implementations, and DL settings.}
    \label{fig:t1plate}
\end{figure*}

T1 estimates remained close to nominal values across MuPa-ZTE implementations, acquisition durations, and reconstruction settings (Figures~\ref{fig:scatter} and~\ref{fig:t1plate}). T1-ZTE and T2T1-ZTE
provided comparable sphere-mean estimates, whereas T1-ZTE generally showed lower within-sphere variability.

Shortening the acquisition predominantly reduced precision. With DL Off, variability increased at shorter durations, particularly for the longest-T1 spheres. DL On reduced variability across all the conditions. At 2 minutes, T1
precision with DL On was comparable to or better than the 4.5-minute DL Off condition (e.g., $\sigma$ = 38 vs.\ 69\,ms for T1-03 with T2T1-ZTE). At 1 minute, denoising remained beneficial, although residual variability was evident for the longest-T1 spheres ($\sigma > 100$\,ms).

\begin{figure*}[!t]
    \centering
    \includegraphics[width=\textwidth]{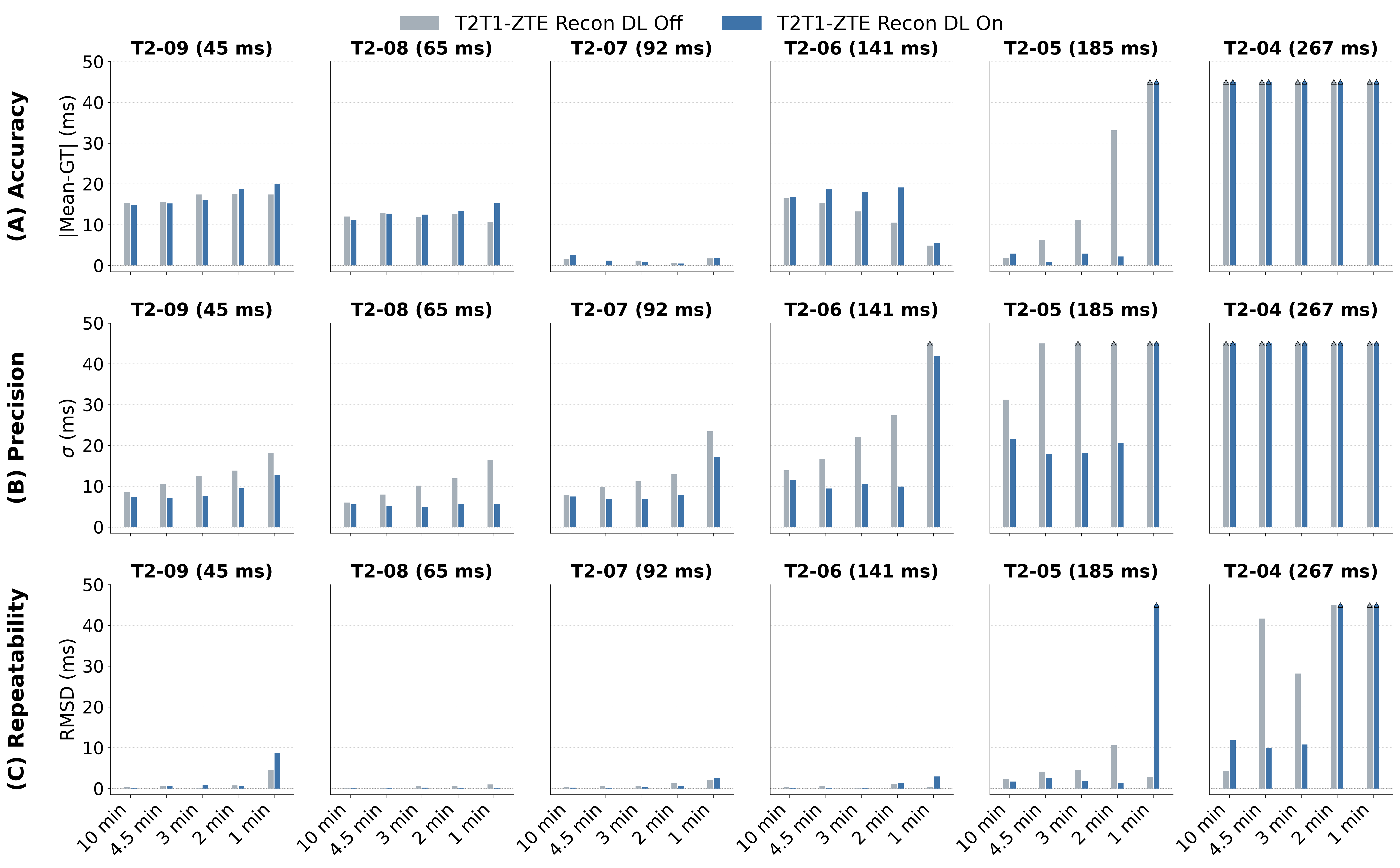}
    \caption{\textbf{T2 accuracy, precision, and short-term repeatability.}
    Sphere-wise \textbf{A}, absolute error relative to nominal T2;
    \textbf{B}, within-sphere standard deviation; and \textbf{C}, absolute
    scan--rescan difference are shown across acquisition durations and DL
    settings.}
    \label{fig:t2plate}
\end{figure*}

T2 performance depended more strongly on relaxation time (Figures~\ref{fig:scatter} and~\ref{fig:t2plate}). Agreement and within-sphere variability remained stable for nominal T2 values of 45.3--141\,ms (absolute error $\leq$20\,ms; $\sigma \leq$10\,ms with DL On down to 2 minutes). The 185-ms sphere remained within 3\,ms of nominal with DL On down to 2 minutes (33\,ms with DL Off at 2 minutes) but showed marked overestimation and increased variability at 1 minute. The 267-ms sphere was consistently overestimated ($>$45\,ms), including at 10 minutes. This duration-independent error indicates limited sensitivity of the implemented T2 encoding rather than an effect of undersampling or denoising.

T1 and brain-relevant T2 estimates remained repeatable down to 2 minutes (scan--rescan RMSD $\leq$2.5\,\% of nominal for T1 and $\leq$6\,\% for T2). Denoising reduced scan--rescan differences for T2 and most T1 spheres, but not for the longest-T1 sphere with T2T1-ZTE. Repeatability deteriorated primarily at 1 minute for long-T1 and longer-T2 spheres (up to 54\,ms for T1-03 and more than 45\,ms for T2-05).

\subsection{Healthy volunteers}
\label{sec:res_volunteers}

\begin{figure*}[htb]
    \centering
    \includegraphics[width=\linewidth]{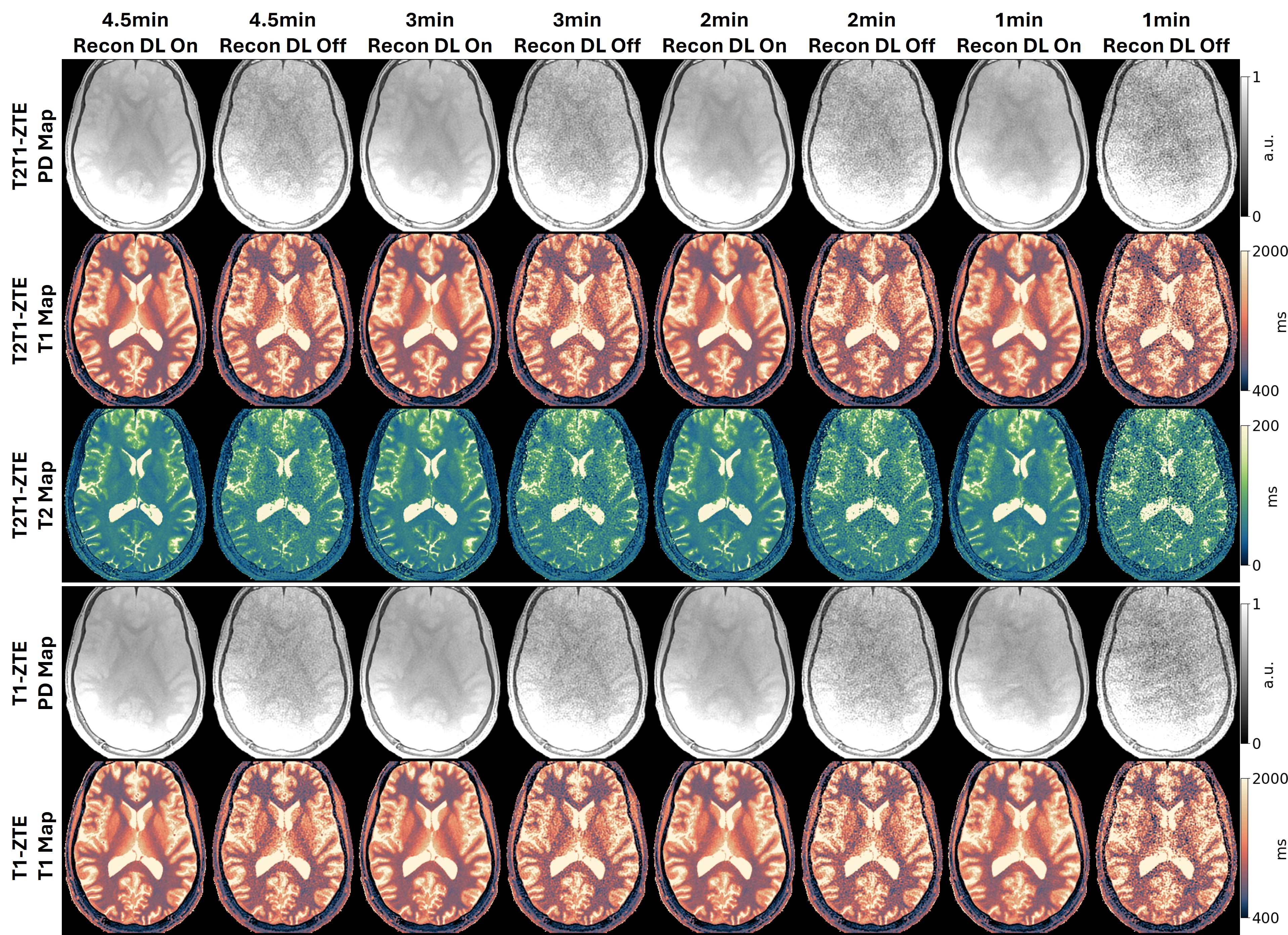}
    \caption{\textbf{MuPa-ZTE maps from Volunteer 1.} Columns show the full
    4.5-minute and retrospectively truncated 3-, 2-, and 1-minute datasets
    with DL On or DL Off. Rows show apparent proton-density, T1, and T2 maps
    from T2T1-ZTE, followed by apparent proton-density and T1 maps from
    T1-ZTE. Display ranges are identical within each parameter.}
    \label{fig:volunteer_maps}
\end{figure*}

\begin{figure*}[htb]
    \centering
    \includegraphics[width=\linewidth]{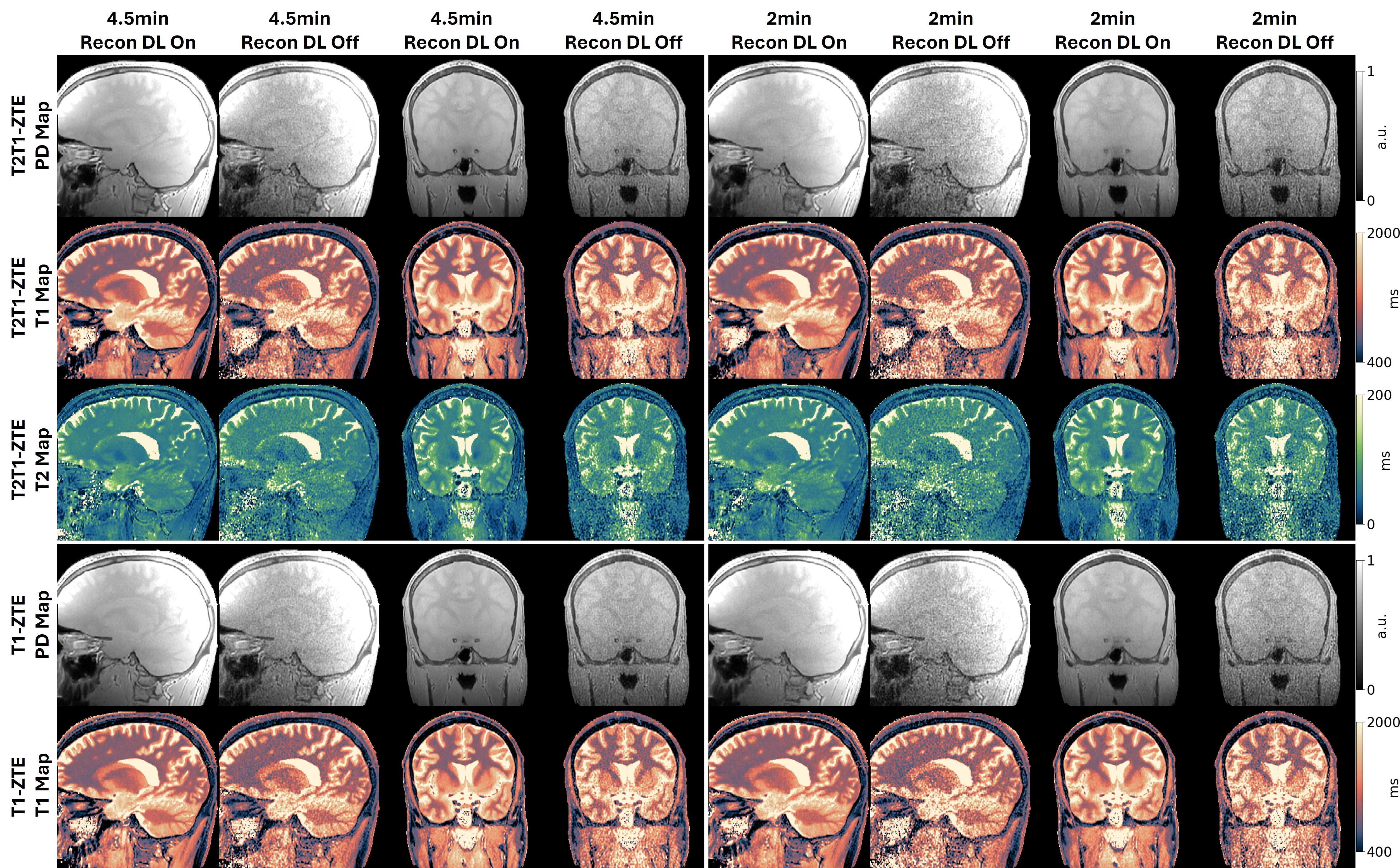}
    \caption{\textbf{Multiplanar MuPa-ZTE maps from Volunteer 1.} Axial,
    sagittal, and coronal reformats are shown for the full 4.5-minute and
    retrospectively truncated 2-minute acquisitions with DL On or DL Off. The isotropic acquisition
    provides consistent anatomical depiction across orientations.}
    \label{fig:volunteer_other_views}
\end{figure*}

Both MuPa-ZTE implementations provided spatially consistent apparent
proton-density and T1 maps, and T2T1-ZTE additionally provided T2 maps
(Figure~\ref{fig:volunteer_maps}). With DL Off, noise increased progressively
at shorter durations. DL On reduced image noise and anatomical
visibility, with the greatest benefit at 2 and 1 minute.
Figure~\ref{fig:volunteer_other_views} shows sagittal and coronal reformats of the maps from the 4.5- and 2-minute acquisitions. With DL On, the 2-minute maps showed clear anatomical detail and approached the overall visual quality of the corresponding 4.5-minute DL On maps.

\begin{figure*}[htb]
    \centering
    \includegraphics[width=\linewidth]{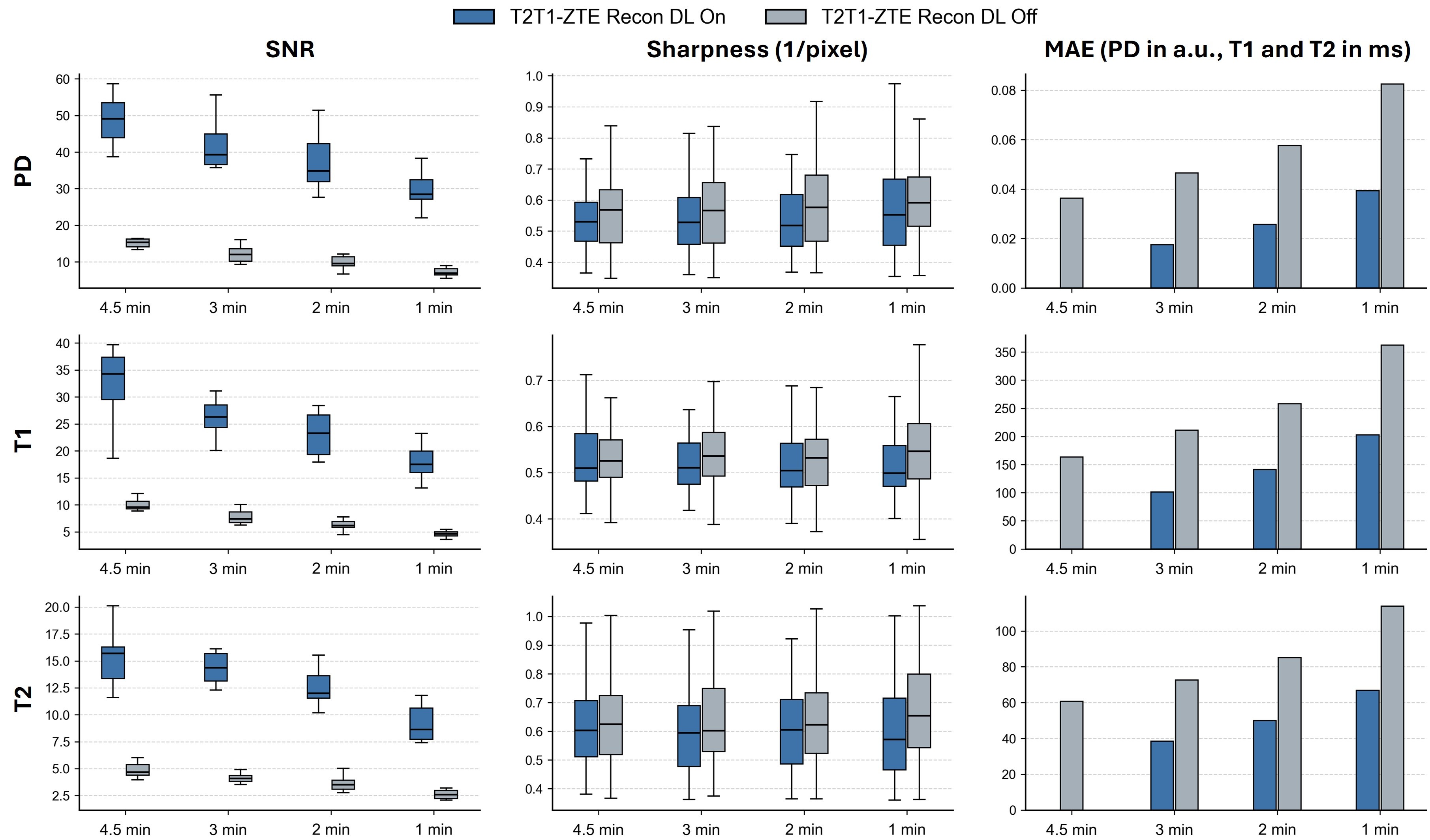}
    \caption{\textbf{Quantitative image-quality comparison of T2T1-ZTE maps.}
    Apparent SNR, edge sharpness, and MAE are shown across acquisition
    durations and DL settings. MAE was calculated relative to the
    corresponding 4.5-minute DL On map and measures consistency with that
    reference rather than absolute quantitative accuracy.}
    \label{fig:image_metrics}
\end{figure*}

Apparent SNR decreased at shorter acquisition durations and increased with DL
On for all three parameters (Figure~\ref{fig:image_metrics}). Edge-sharpness
distributions were comparable between DL Off and DL On, without consistent
evidence of edge broadening.

MAE relative to the 4.5-minute DL On reference increased at shorter durations
but was generally lower with DL On. The 2-minute DL On maps had lower MAE than
the 4.5-minute DL Off maps. Because the reference was processed with DL On and
shared acquired data with the truncated conditions, this finding indicates
consistency with the selected reference rather than independent accuracy.
Volunteer 2 showed the same overall behavior (Supporting Information~B, Figure~S3). 
Denoising reduced voxel-wise scan--rescan differences, with residual differences most visible in CSF, gray matter, and tissue boundaries (Supporting Information~B, Figures~S4 and~S5).

\begin{figure}[htb]
    \centering
    \includegraphics[width=\linewidth]{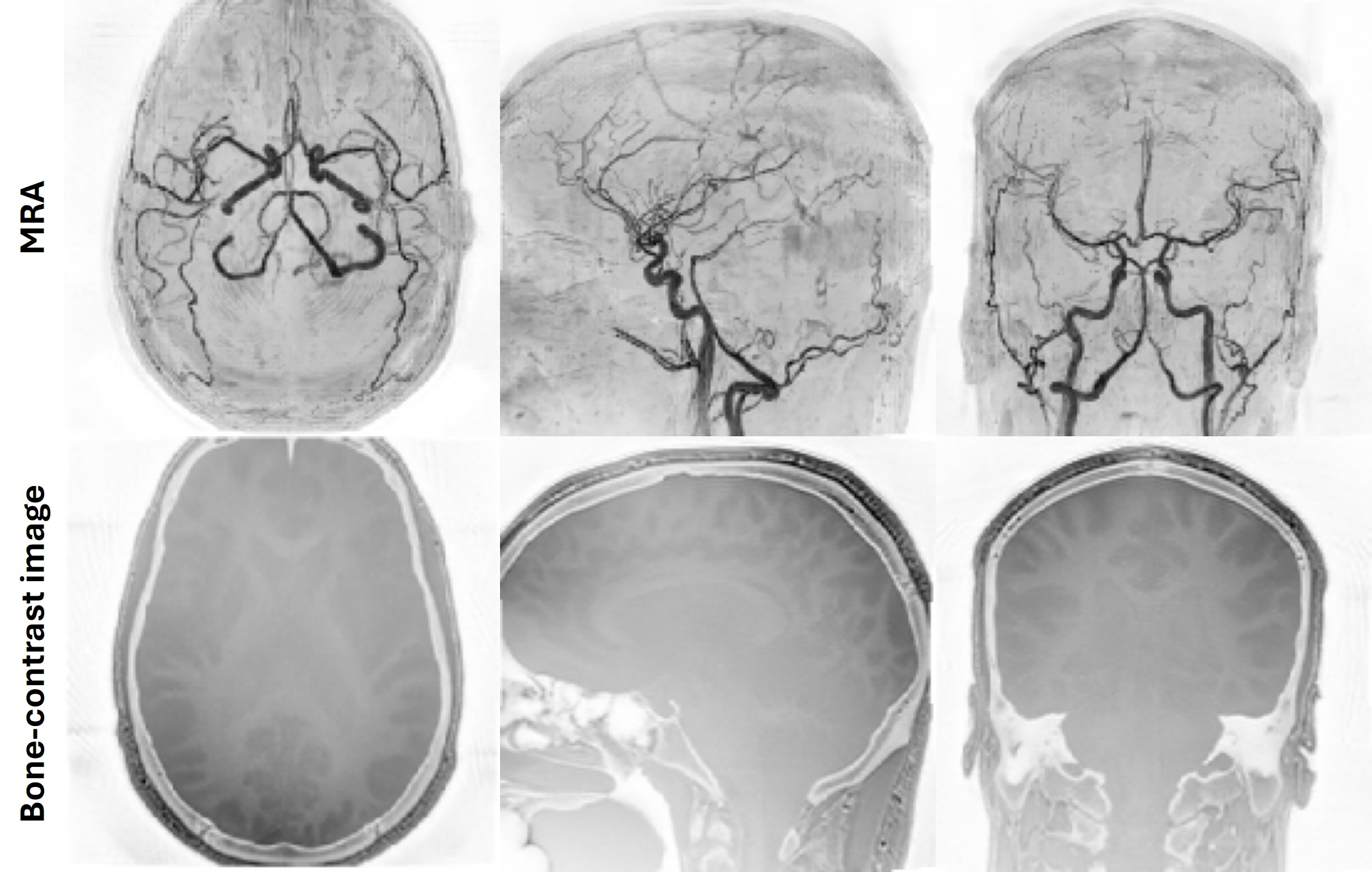}
    \caption{\textbf{Complementary contrasts derived from MuPa-ZTE without
    additional acquisition time.} Minimum-intensity projection of a
    post-inversion T1-ZTE source image ($S_2$) shows inflow-sensitive angiographic
    contrast. Intensity inversion of native ZTE ($S_0$) provides CT-like cranial-bone
    contrast. Both images were derived from the full 4.5-minute acquisition
    with DL On.}
    \label{fig:additional_features}
\end{figure}

Intensity-inverted native ZTE provided CT-like cranial-bone contrast, while a
post-inversion T1-ZTE image provided inflow-sensitive angiographic contrast
using minimum-intensity projection
(Figure~\ref{fig:additional_features}).

\subsection{Patient demonstration}
\label{sec:res_patient}

\begin{figure*}[htb]
    \centering
    \includegraphics[width=\linewidth]{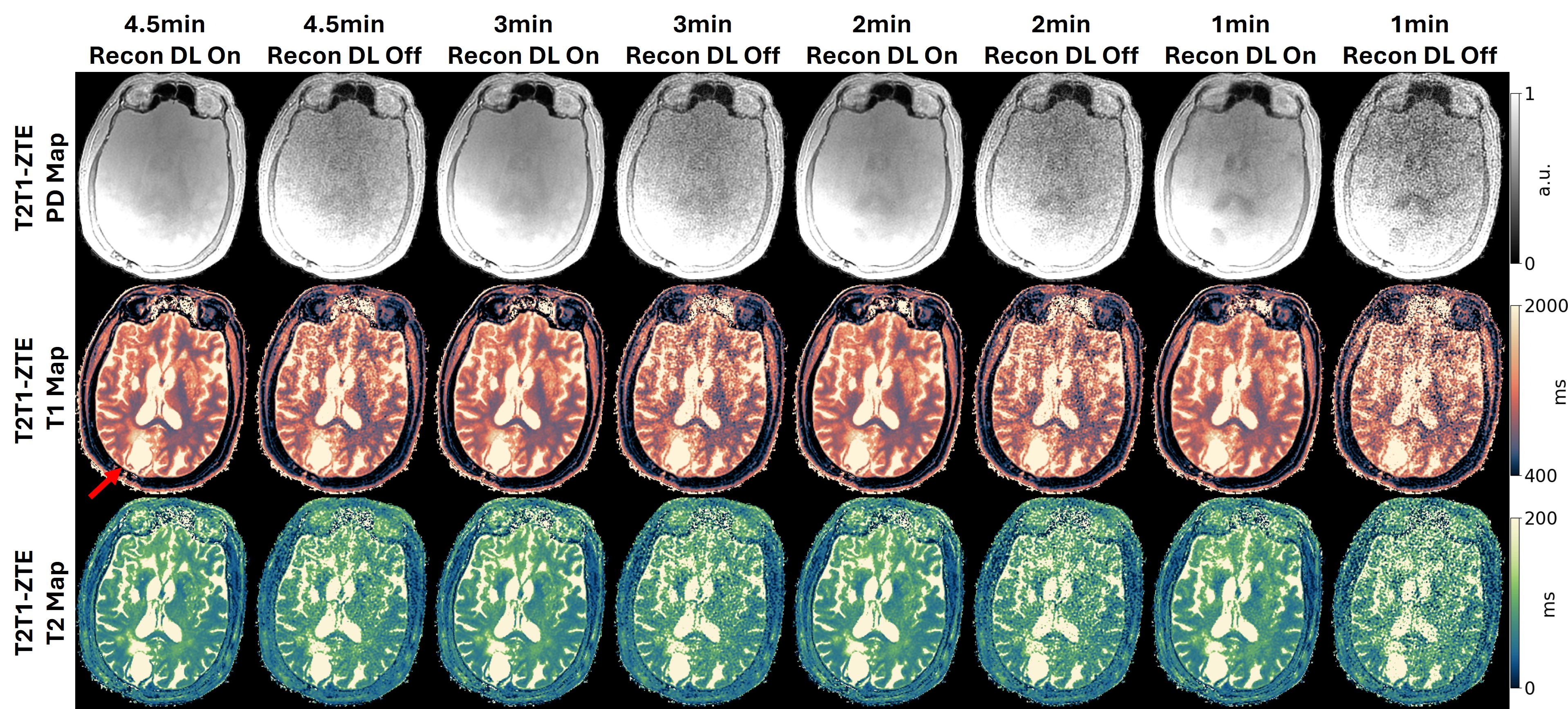}
    \caption{\textbf{T2T1-ZTE in a patient with brain metastases.} Columns
    show the full 4.5-minute and retrospectively truncated 3-, 2-, and
    1-minute datasets with DL On or DL Off. Rows show apparent
    proton-density, T1, and T2 maps. The arrow indicates the known lesion.}
    \label{fig:patient_maps}
\end{figure*}

The 4.5-minute T2T1-ZTE acquisition provided whole-brain apparent
proton-density, T1, and T2 maps in which the known lesion and surrounding
anatomy were visible (Figure~\ref{fig:patient_maps}). At 2 minutes with DL On,
the lesion and major anatomical structures remained visible. At 1 minute,
noise and spatial variability increased, particularly in T2.

This single case demonstrates technical feasibility but does not establish
diagnostic performance.

\section{Discussion}
\label{sec:discussion}

MuPa-ZTE combines steady-state native ZTE with transient-state
magnetization-prepared ZTE for quiet, rapid, isotropic 3D multiparametric
mapping. Native ZTE provides apparent proton density and an independent
normalization reference, while the magnetization-prepared ZTE segments generate images varying in T1 and T2 weighting. The evaluated 4.5-minute T2T1-ZTE implementation provided robust T1
mapping and T2 mapping over a range relevant to brain parenchyma. Deep
learning-based denoising reduced variability and provided promising
retrospective results at 2 minutes.

\subsection{MuPa-ZTE as a multiparametric extension of ZTE}
\label{sec:disc_extension}

MuPa-ZTE extends the established ZTE acquisition platform from predominantly
proton-density-weighted imaging to multiparametric mapping. It combines two
complementary signal regimes within a common 3D radial architecture:
steady-state native ZTE provides apparent proton-density information, while
transient-state ZTE following T1 or T2 preparation provides relaxation
sensitivity. This integration preserves characteristic ZTE properties,
including short TR, high acquisition duty cycle, isotropic center-out
sampling, low gradient switching, and sensitivity to rapidly decaying signals.

Conceptually, MuPa-ZTE is related to magnetization-prepared FLASH and 3D-QALAS
but replaces the conventional segmented spoiled gradient-echo readout with
ZTE.\cite{bib_b25,bib_b26,bib_b27,bib_b47} Native ZTE also provides an independent
amplitude reference, enabling least-squares dictionary matching based on both
the shape and relative amplitude of the prepared signal evolution. MuPa-ZTE
thereby combines efficient transient-state encoding with the distinctive
acquisition properties of ZTE.

Matched phyllotaxis sampling further integrates the contrast states. Each
acquisition repetition contributes the same angular sampling pattern to all
source images, minimizing sampling-dependent differences among contrasts.
Retaining complete repetitions allows the acquisition duration to be adjusted
without changing the contrast-encoding structure. The trajectory is also
compatible with self-navigation and retrospective motion
correction.\cite{bib_b23}

The framework is modular rather than limited to the two implementations
evaluated here. T1-ZTE provides a focused option when T2 mapping is not
required, whereas T2T1-ZTE provides apparent proton density, T1, and T2
information within one acquisition. Other preparation modules could target
diffusion, magnetization transfer, chemical exchange, or selective tissue
suppression while retaining the common ZTE readout and sampling architecture.
Each extension would require application-specific optimization and validation.
MuPa-ZTE therefore provides a flexible framework for extending ZTE toward
quantitative multiparametric imaging.

\subsection{Quantitative performance}
\label{sec:disc_quantitative}

T1 estimates were stable across implementations and acquisition durations.
Shortening scan time primarily affected precision rather than sphere-mean agreement.
T1-ZTE generally showed lower variability and smaller scan--rescan
differences than T2T1-ZTE, consistent with its focused T1 encoding.

T2 sensitivity was more range dependent than T1. A single T2-preparation echo time of
80\,ms provided stable estimates over approximately 45--141\,ms, and the
185-ms sphere remained reliable through 2 minutes. The 267-ms sphere was
overestimated even at full duration, indicating insufficient long-T2
discrimination.

The current implementation is therefore most sensitive to T2 values
characteristic of brain parenchyma. Longer or multiple T2-preparation times
could improve long-T2 sensitivity\cite{bib_b48}, whereas shorter preparation times may
improve encoding of shorter-T2 tissues. This represents an optimization
opportunity rather than a fundamental limitation of MuPa-ZTE.

The phantom measurements should not be interpreted as establishing absolute in
vivo accuracy.\cite{bib_b49} Effective RF-efficiency factors were selected using the same
full-duration phantom dataset used for evaluation, potentially producing
optimistic agreement. Independent calibration, B1 mapping, and comparison with
established in vivo reference methods are required.

\subsection{Apparent proton density and dictionary matching}
\label{sec:disc_pd}

Native ZTE provides both an apparent proton-density image and the reference
for normalizing prepared signals. This reduces the parameter search to T1 and
T2 for T2T1-ZTE and T1 for T1-ZTE.

Unlike normalized inner-product matching, least-squares matching retains the
amplitudes of the prepared signals relative to $S_0$. This provides additional
discrimination among parameter combinations that produce similar
signal-vector shapes and is particularly useful for the compact set of
contrast states used by MuPa-ZTE.

The native image is described as apparent proton density because no absolute
calibration was performed. It also retains residual relaxation dependence,
system scaling, and sensitivity to rapidly decaying components.

\subsection{Quiet ZTE neuroimaging}
\label{sec:disc_quiet}

Low acoustic noise is a defining property of ZTE. Conventional MRI acoustic
noise arises primarily from gradient-coil vibration during rapid changes in
gradient current. ZTE reduces this source by maintaining the readout gradient
and applying small directional updates between
repetitions.\cite{bib_b4,bib_b6} MuPa-ZTE retains this gradient architecture
across all contrast states.

Quiet mapping may be particularly useful in pediatric, older, anxious, or
acoustic noise-sensitive populations and in repeated examinations. Low acoustic noise may also reduce confounding in functional and auditory
neuroimaging.\cite{bib_b4}

The present study did not include a sequence-specific acoustic comparison. The
characterization of MuPa-ZTE as quiet is based on preservation of the
established ZTE gradient architecture. Direct acoustic comparison with
conventional multiparametric protocols remains warranted.

\subsection{Acquisition duration and reconstruction}
\label{sec:disc_duration}

Conventional reconstruction provided whole-brain 1.1-mm isotropic maps in 4.5
minutes. Denoising increased apparent SNR and reduced variability without
consistent loss of measured edge sharpness. The 2-minute DL On condition
retained encouraging anatomical and quantitative quality.

Denoising cannot replace parameter encoding. It reduced variability but did
not correct the long-T2 limitation. MuPa-ZTE also does not depend on deep
learning processing. Conventional reconstruction establishes the acquisition
and mapping framework, whereas denoising provides one route toward shorter
scans.

Joint reconstruction across contrast states, locally low-rank methods, or
subspace reconstruction may provide further gains. Previous silent
quantitative ZTE work has demonstrated reconstruction using Bloch-simulated
temporal subspaces.\cite{bib_b50}

\subsection{Complementary contrasts and applications}
\label{sec:disc_contrasts}

MuPa-ZTE source images provide information beyond the parameter maps. Native
ZTE supports cortical-bone visualization and synthetic CT
generation.\cite{bib_b7,bib_b8,bib_b9,bib_b10,bib_b11,bib_b51}
Post-inversion images provide anatomical and inflow-sensitive vascular
contrast. These outputs were obtained without separate acquisitions but were
evaluated only qualitatively.

Nominal $TE = 0$ retains sensitivity to rapidly decaying components that are
not captured by conventional echo-based
relaxometry.\cite{bib_b4,bib_b5,bib_b15} The apparent proton-density, T1, and
T2 estimates may consequently contain contributions from multiple tissue
compartments. Applications involving myelin, bone, collagen, or other short-T2
tissues will require appropriate multicomponent signal models.

The patient case provides an initial demonstration of technical feasibility.
Larger studies with independent reference imaging are required to determine
diagnostic performance and clinical utility.

\subsection{Limitations and future work}
\label{sec:disc_limitations}

The in vivo evaluation was limited to two healthy volunteers and one patient.
Shorter durations were generated retrospectively, no independent in vivo
relaxometry reference was acquired, and scan--rescan measurements were
performed without repositioning.

Fixed RF-efficiency factors and the segment-averaged signal model may not
fully represent spatially varying preparation effects or spoke-wise
magnetization evolution. Off-resonance, magnetization transfer, chemical
exchange, multicomponent relaxation, and residual transverse coherence were
not modeled explicitly.

Future work should prioritize prospective evaluation of the 2-minute protocol,
independent calibration, B1 correction, comparison with established
relaxometry, intersession repeatability, multicenter phantom studies, motion
correction, and clinical validation. Application-specific preparation timing
and joint reconstruction may further improve parameter sensitivity and
acquisition efficiency.

\section{Conclusion}
\label{sec:conclusion}

MuPa-ZTE provides a flexible framework for quiet, isotropic 3D
multiparametric mapping by combining native and magnetization-prepared ZTE
within a common acquisition architecture. The evaluated 4.5-minute
implementation provided apparent proton density, robust T1 mapping, and T2
mapping over a range relevant to brain parenchyma while retaining the nominal
zero echo time, efficient center-out sampling, low gradient-switching
characteristics, and sensitivity to rapidly decaying signals of ZTE.
Deep learning-based denoising reduced variability and supported promising
retrospective results at 2 minutes. Beyond the parameter maps, the same
acquisition provided complementary bone-sensitive and inflow-angiographic
contrasts. MuPa-ZTE therefore extends the established advantages of ZTE into
multiparametric relaxometry and provides a foundation for independent
validation, application-specific optimization, and broader clinical
investigation.

\section*{Acknowledgments}
This publication is part of the ``Trustworthy AI for MRI" ICAI lab within the project ROBUST: Trustworthy AI-based Systems for Sustainable Growth with project number KICH3.LTP.20.006, financed by the Dutch Research Council (NWO), GE HealthCare, and the Dutch Ministry of Economic Affairs and Climate Policy (EZK) under the program LTP KIC 2020-2023.
During manuscript preparation, ChatGPT 5.6 was used to assist with textual
editing, including improvements to language, clarity, concision, and
organization. The authors critically reviewed and verified all edited text,
references, scientific statements, and conclusions and take full
responsibility for the content of the manuscript.

\section*{Author Contributions}
\textbf{Alireza Samadifardheris:} Conceptualization; Data curation; Formal analysis; Investigation; Methodology; Software; Validation; Visualization; Writing -- original draft; Writing -- review and editing.
\textbf{Shishuai Wang:} Conceptualization; Data curation; Formal analysis; Investigation; Methodology; Software; Validation; Visualization; Writing -- original draft; Writing -- review and editing.
\textbf{Ana Beatriz Solana:} Methodology; Writing -- review and editing.
\textbf{Jose de Arcos:} Resources; Software; Writing -- review and editing;
\textbf{Noemi Sgambelluri:} Data curation.
\textbf{Emil Langensee Ljungberg:} Conceptualization; Methodology; Writing -- review and editing.
\textbf{Sagar Mandava:} Software; Writing -- review and editing.
\textbf{Mika Vogel:} Project administration.
\textbf{Steven C. R. Williams:} Writing -- review and editing.
\textbf{Stefan Klein:} Funding acquisition; Supervision; Validation; Writing -- review and editing.
\textbf{Juan Antonio Hernandez-Tamames:} Funding acquisition; Project administration; Resources; Supervision; Validation; Writing -- review and editing.
\textbf{Dirk H. J. Poot:} Conceptualization; Funding acquisition; Methodology; Project administration; Resources; Software; Supervision; Validation; Writing -- review and editing.
\textbf{Florian Wiesinger:} Conceptualization; Funding acquisition; Methodology; Project administration; Resources; Software; Supervision; Validation; Writing -- review and editing.

\section*{Data Availability Statement}
The human imaging data are not publicly available because of
participant-privacy and institutional restrictions. Deidentified data or
representative datasets may be made available by the corresponding author upon
reasonable request, subject to ethics approval and applicable institutional
agreements.
A research work-in-progress implementation of MuPa-ZTE may be made available
to eligible academic collaborators for evaluation on compatible GE HealthCare
MR systems, subject to applicable institutional, contractual, safety, and
regulatory requirements.

\end{document}